# THE PHENOMENAL UNIVERSE

## Copyright 2009 by Paul E. Field, Ph.D.[1]

A wave model for the interconversion of photon particle behavior and wave phenomena is examined in relation to the adiabatic spatial expansion of the Universe. An historical perspective of the mathematics of physical science over the 20th century between what are termed phenomenal and noumenal are contrasted based on quotations from leading scholars. The question is posed whether the generation of space is an energetic sink and should be included in the conservation of energy as posited by the First Law of Thermodynamics. Various aspects of space expansion and electromagnetic radiation are examined as well as a mechanism based on imaginary conmponents to real space dimensions and the consequent creation of the time dimension.

As a physical chemistry student in the late '50s studying thermodynamics I was taught in a tradition that emphasized the phenomenological nature of the subject. Although I appreciated the nature of the approach I always felt a reservation with the term even though I used it when necessary - it just seemed to be too much of a mouthful. Now at the end of my career and choosing a title for this essay, I find there is a better word with fewer syllables and a powerful double connotation. I have also discovered that the root noun has had its definition completely rewritten! So it seems the change in terms is overdue. (Italics are from the earlier edition.)

From Webster's New Collegiate Dictionary, G. & C. Merriam Co., Springfield, MA 1980, cf. *1951*:
**phenomenology** (cf. *n. 1. The branch of a science dealing with the description and classification of phenomena. 2. Scientific description of actual phenomena, with avoidance of all interpretation, explanation, and evaluation.*) *n.* **1 :** the study of development of human consciousness and self-awareness as a preface to philosophy or a part of philosophy **2 a** (1) : the description of the formal structure of the objects of awareness and of awareness itself in abstraction from any claims concerning existence [the ~ of internal time-consciousness] (2) : the typological classification of a class of phenomena [the ~ of religion] **b :** an analysis produced by phenomenological investigation
**phenomenological** *adj.* **1 :** of or relating to phenomenology **2 :** PHENOMENAL **3 :** of or relating to phenomenalism
**phenomenal** *adj.* : relating to or being a phenomenon; as: **a:** known through the senses rather than through thought or intuition. **b:** Concerned with phenomena (cf. *observed data*) rather than with hypothesis, (cf. *as phenomenal science*) **c:** EXTRAORDINARY, REMARKABLE **syn.** see MATERIAL **ant.** noumenal

The nineteenth century witnessed the culmination of *phenomenal* physics as mechanics, electromagnetics, fluid mechanics, and thermodynamics were elaborated with mathematical finesse. It was with the dawn of the twentieth century that physics divided into two branches to deal with the microscopic (quantum mechanics) and the macroscopic (relativity theory and cosmology). Although some crossover developed by the end of the twentieth century (e.g. quantum gravity) the division was practically a schism. Interestingly enough, one gets the feeling that even the mathematics used in the elaboration of quantum phenomena developed a sort of schizophrenia with the different approaches of wave mechanics and matrix mechanics. Both of course were shown to be equivalent but their separate approaches were singularly different. Schroedinger's wave mechanics was rooted in what might even be called *phenomenal* math while Heisenberg's mechanics never made claims of physical interpretation and might even be labeled *noumenal* math. Further, as the 20th century evolved the theoretical physicists continued down Heisenberg's path employing more and more exotic algebras and calculli borrowed from the mathematicians; most notably spending the last thirty years inventing string theory without a single testable hypothesis according to Woit (1) and Smolen (2). Symptomatic of this approach is Dirac's statement (3): "Einstein has stated that he was indebted to Mach's principle . . . which led him to general relativity. I do not see how this can be. . . . One should keep the need for a sound mathematical basis dominating one's search for a new theory. Any physical or philosophical ideas that one has must be adjusted to fit the mathematics. Not the other way around."

There are, of course, a myriad of unique mathematical relationships with some of them so broadly applicable to physical phenomena that they all might be called *phenomenal* math. Dirac's position notwithstanding, it is even more important to acknowledge that the theories of science must first be on phenomena and not on the mathematics. G. N. Lewis described the period during the first quarter of the twentieth century prior to the quantum break-out as the beginning of the third stage of thermodynamics following first the development of basic principles of energy (e.g. Joule and Carnot) in the first half of the 19th century and then by the building up of theorems (e.g. J. W. Gibbs and Kelvin) in the second half. The third stage was seen as the development of methods and applications and the accumulation of data but to be carried on in the established tradition. The 1923 publication of the treatise by G. N. Lewis and its 1961 revision reflected this tradition (4). Lewis wrote: "The formal severity of mathematics has its disadvantages. Indeed in our opinion absolute mathematical rigor is a sort of *ignis fatuus* (foolish beacon or will-o'-the-wisp), which must not serve as a guide to the scientific investigator, although we do not claim that its pursuit, with proper safeguards, may not offer a very wholesome exercise." Lewis's students and colleagues perpetuated this tradition. Tolman states (5) "In the presentation of the material, the endeavour will be made to emphasize the physical nature of assumptions and conclusions and their interconnexion, rather than to lay stress on mathematical generality or even, indeed, on mathematical rigour. . . . Throughout the essay a macroscopic and *phenomenological* point of view will be adopted as far as feasible." Finally, the following quotations by renowned scientists-mathematicians-philosophers (each) were cited by physical chemist Joel H. Hildebrand (6):

G. N. Lewis: "Mathematics sometimes serves as a substitute for thought." (recollected by JHH)
J. Schwartz: "The simple-mindedness of Mathematics:-its willingness, like that of a computing machine, to elaborate upon any idea, however absurd; to dress scientific absurdities alike in the impressive uniform of formulae and theorems. Unfortunately, however, an absurdity in uniform is far more impressive than an absurdity unclad." (7)

A. N. Whitehead: "There can be no true physical science which looks to mathematics for the provision of a conceptual model. Such a procedure is to repeat the errors of the logicians of the Middle Ages."

The history of science is filled with instances when some branch of a scientific discipline found the mathematics already invented that could be applied to the problem at hand as was the case with Riemann's curved surfaces and spaces adopted by Einstein's general field theory. But well known exceptions include Newton's invention of his differential calculus (although he refrained from its use in the Principia) and Minkowski's development of hyperbolic space to place Lorentz and Einstein's special relativity on firmer ground. It is enlightening reading as Minkowski bemoans the fact that pure mathematics missed calling into question the Newtonian reliance on absolute space. The pursuit of *noumenal* mathematical physics such as string theory and quantum gravity, for example, are not being questioned, they are without doubt "wholesome exercises" that may yet yield predictable *phenomena* to be incorporated into the body of experimentally verifiable physics. This essay is an attempt to examine alternative interpretations of some *phenomenal* aspects of space, time, energy and matter, and in particular to suggest that the First Law of Thermodynamics needs to be amended. The present epoch of the Universe began after the first half million years following the Big Bang when atoms formed and light is said to have decoupled from matter. Although some speculation of the earlier epochs will be made because of its mathematical suggestiveness, this discussion will be essentially limited to this epoch because the same phenomena are on-going.

## LIGHT

Light, i.e. electromagnetic radiation, is a quantized wave phenomena first rationalized by Planck in 1900 when he showed that light energy is quantized ($E = h\nu$) to explain black body radiation. It was subsequently shown to exist as discrete particles (called corpuscles by Newton) called photons by Einstein in 1905 to explain the photo-electric effect. One could imagine that the photon exists as a discrete single wave (wavelet) such as the first Hermite polynomial solution for a one dimensional harmonic oscillator: HO given by the equation: $\Psi = x \exp(-\alpha x^2)$.

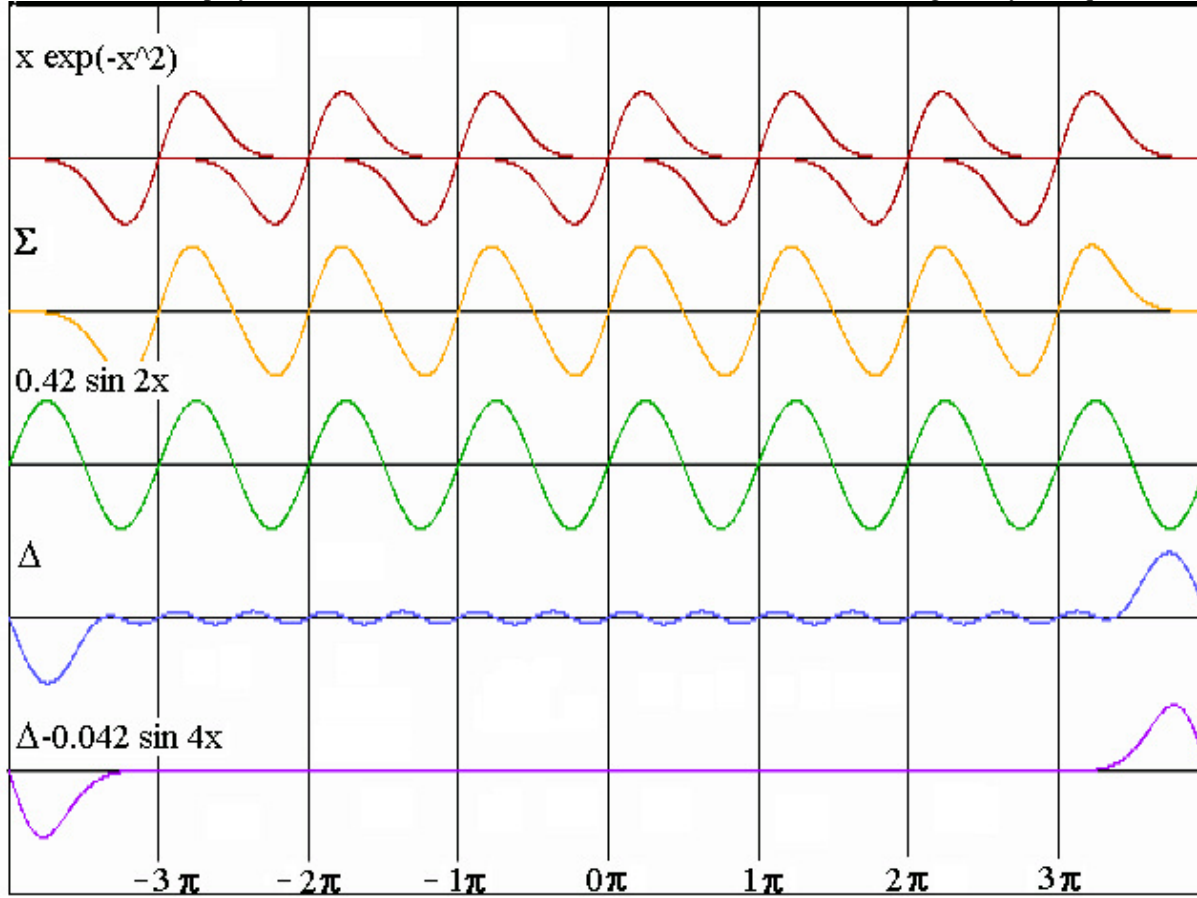

As shown in Figure 1, individual wavelets can add in phase to form successively longer discrete sinusoidal wavetrains. When individual HOs are displaced by a distance $\pi$ and added together, seven such HOs shown in the figure between $+4\pi$ and $-4\pi$ the resultant wave is sinusoidal. A matching sine wave of two times the frequency of the summed HOs with an amplitude of about 42% leaves a difference of doubled frequency, i.e. 4 times, having an amplitude about one-tenth the original sine wave and a cancellation of the signal on subtraction leaving a residual smaller than the resolution of the graph (say 1 part in 350) except at its ends which appear to form the two halves of the component exponential function. The equivalence seems reasonable given the series expansions for the functions(8)[2]:

$$\exp(x) = e^x = 1 + x + (x^2/2!) + (x^3/3!) + \ldots$$

substituting $-x^2$ for x and premultiplying by x yields:

$$x \exp(-x^2) = x - x^3 + (x^5/2!) - (x^7/3!) + \ldots$$

compares by terms to:

$$\sin x = x - (x^3/3!) + (x^5/5!) - (x^7/7!) + \ldots$$

substituting 2x for x yields:

$$\sin 2x = 2x - (2^3 x^3/3!) + (2^5 x^5/5!) - (2^7 x^7/7!) + \ldots$$

A second property of these functions is the area bounded by the individual maxima. The integral from zero to infinity of the HO is $1/2\alpha$ or with $\alpha = 1$ is 0.5 so that the absolute value of both peaks is normalized, i.e. equal to one. The integral of sin x between 0 and $\pi$ is 2. Of course, the integrals over both waves for a full cycle sums to zero. It should be noted that setting $\alpha = 0.5$ to equate the areas of the two functions doubles the width of each HO and makes the trough and crest of successive HOs overlap with the summed wave becoming very complex.

Given that the photon has a finite velocity it's one dimension has a preferred direction in spacetime. The usual formulation for a traveling electromagnetic wave is given in terms of E and B the electric and magnetic field strengths respectively as: $E = E_m \sin(kx-wt)$ and $B = B_m \sin(kx - wt)$, where k has the units radian/cm and w the units radian/sec and the speed of light $c = w/k$. The substitution of the quantity in parentheses, (kx-wt), for x in the HO equation defines the photon in terms of the eletromagnetic field properties. Its finite width would be dependent on the uncertainty principle but clearly a wavelength orders of magnitude greater than its width (cross-section diameter, much like a spaghetti). As the HO wave can form in-phase trains (lengthen) it is reasonable that it can also form in-phase bands (widen - like fettuccini or large bundles of spaghetti) in order to act as Huygen's wavefronts. These enlarged units can be regarded as coherent "packets." The variation in amplitude should not be imagined as a spatial dimension but rather as a spatial intensity whose squared value would be associated with the probablity of locating an individual photon. As the leading edge of a photon has a finite amplitude well in advance of its maxima (presumably to infinity but practically limited by the uncertainty principle) it is quite reasonable to understand how it can anticipate the existence of one or more slits in its path and subsequently scatter from a packet into a rain of individual photons to form interference patterns where two slits exist.

RED-SHIFT

Having a phenomenal model of quantized radiation, the question now arises about the behavior of radiation in the presence of gravitational fields. This phenomenon is associated with the loss of energy of a photon escaping from a graviational field such as from a star. It is witnessed as the lengthening of its wavelength, the so-called red-shift. As noted above, the energy of a photon is proportional to its frequency: $E = h\nu$. It travels at the speed of light c and its wavelength is related to its frequency as $\lambda = c/\nu$. The successive lengthening of the photon's wavelength, taken as an HO wave is illustrated in Figure 2. If the exponential constant, $\alpha$ in the above equation, is included the consequent reduction in area equals $1/2\alpha$. When the wavelength is extended by the same factor to $\alpha x$, the area of the wave's displacement is restored to the original: $\alpha/2\alpha = 0.5$.

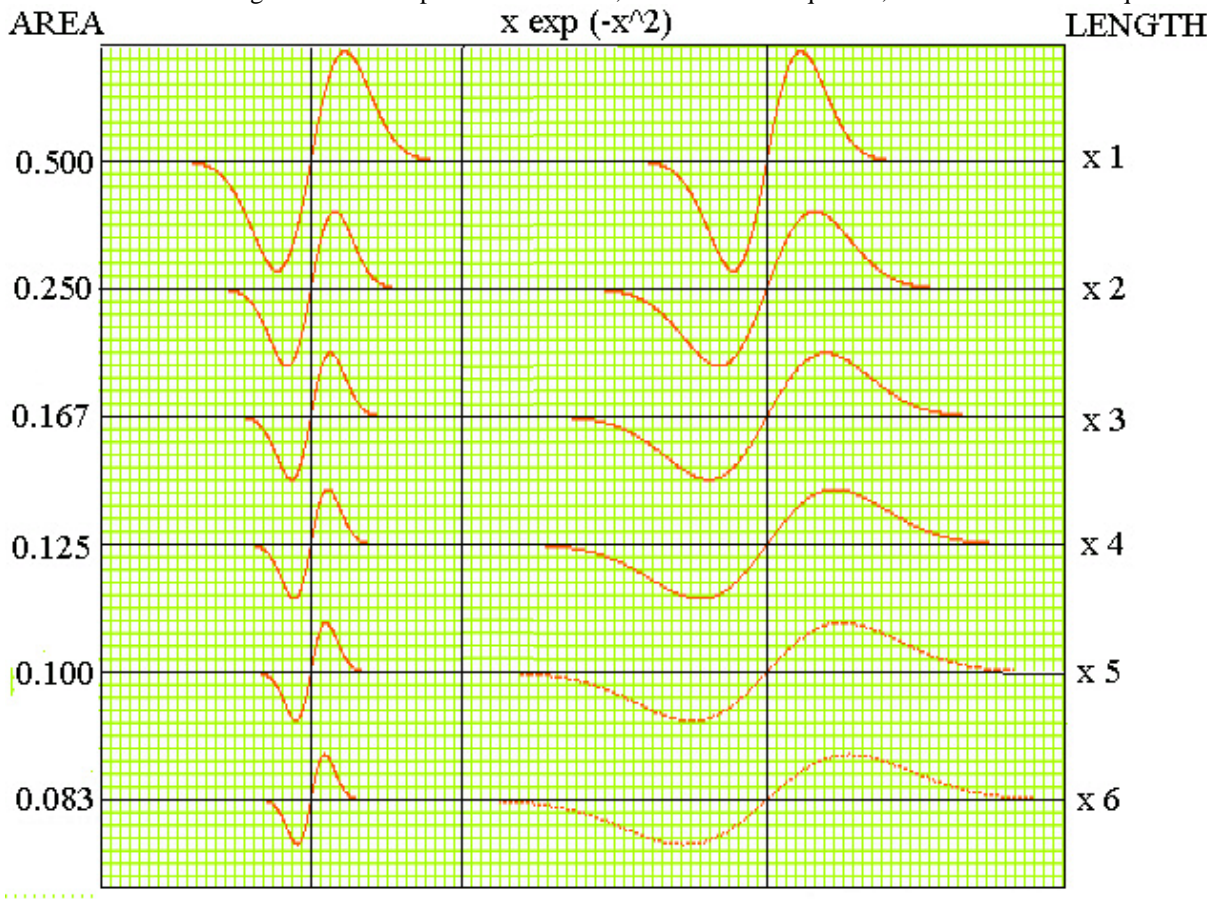

In the figure, there are an estimated 48 blocks in each half of the waves' right hand graphs and 48, 24, 16, and 12 in the first four on the left. The intensity of a wave is known to be proportional to the square of the amplitude of the signal. By squaring the amplitude the negative trough and positive crest both contribute positively. This is the same criterion used in wave mechanics to analyze the probability of the location of a particle, e.g. an electron, based on the square of its amplitude wavefunction. Ordinarily, the integral of any wavefunction, $\psi$, is multiplied by an adjustable constant in order to normalize the value of the integral of $\psi^2$ over all space to unity. This implies that the probablity of locating the particle described by the wavefunction is somewhere in the universe and therefore really exists. This concept of having a probablity of location is as valid for photons as it is for electrons which are either bound in an atom or free like photons. For the area of the red-shifted photons in this phenomenal model to be unity, the normalization constant for the squared functions turns out to be $8(\alpha/2\pi)^{1/2} \sim 3.1915\alpha^{1/2}$. The first six integral lengths and their squares are shown in Figure 3. Numerical integration of the right half of both sets of curves (for $0 < x < 5$) by the computer program that made the plots gave a value for the areas of $0.50000 + 0.00001$ before showing any residual error. Note that the amplitude is clearly smaller than unity while the squared waves are adjusted by the normalization factor. The maximum in the fundamental curve occurs at a distance from its center of $1/2^{1/2} = 2^{1/2}/2 = 0.7071...$ with an amplitude of $0.428883...$. This amplitude is the factor used for the amplitude of the sin wave in comparison with the HO wave in the first figure. It is obtained from the series expansion of $x \exp(-x^2)$ by setting $x=1/2^{1/2}$ and summing the first seven terms (up to the $x^{13}$th term) to get six digit precision.

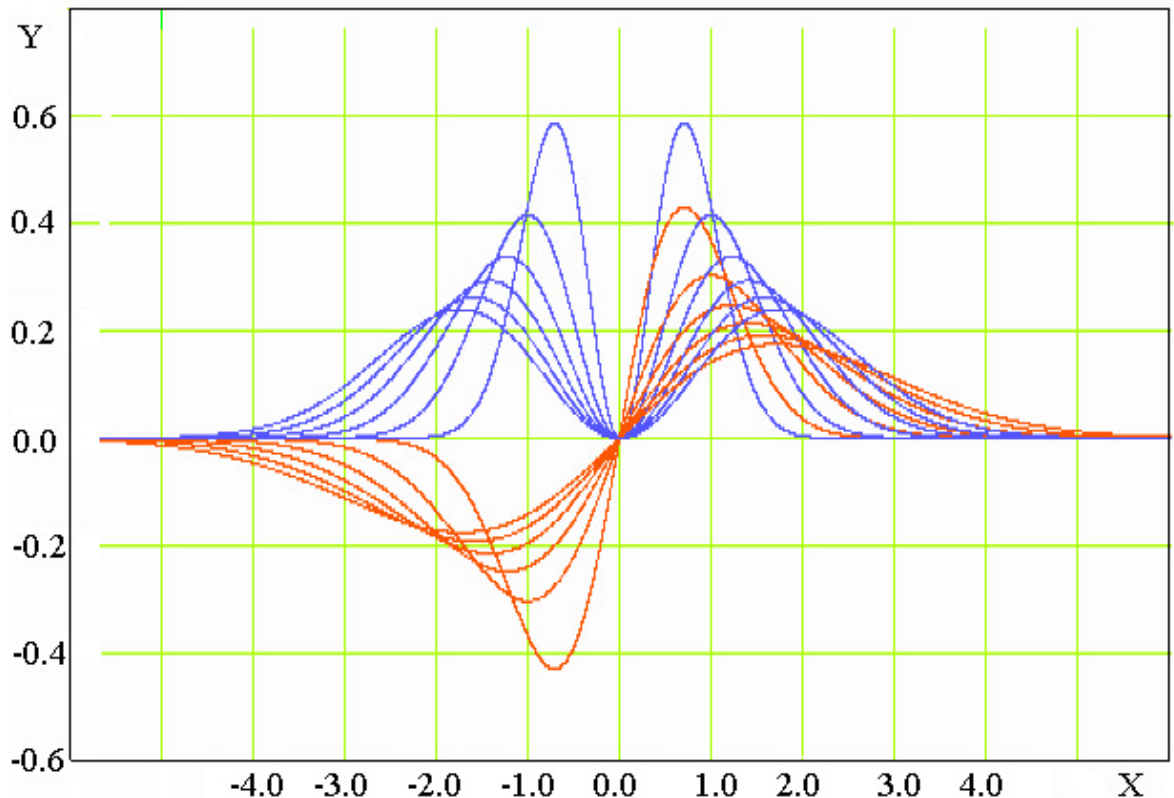

As the area is spread over a longer and longer length and the amplitude (and its square) are proportionally diminished we are forced to the conclusion that in sufficient time and after sufficient red-shift the photon's location can be anywhere along its path. In the classical sense the path of the photon is a ray composed of in phase fluctuating magnetic (B) and electric (E) fields perpendicular to each other along the path. As the probability of the location of the photon spreads out along the path so does the linear extent of these fields. Since the energy of the photon is conserved this phenomenon thus implies that it is stored in the space (vacuum) as what ought to be considered dark energy. Alternatively, since energy and mass are equivalent according to the well known Einstein relation, $E = mc^2$, this dark energy may be presumed to be equivalent to dark matter. It might even be suggested that the radiation is not waves *in* space but *of* space.

EXPANSION AND COMPRESSION

A further consideration in the stretching of the photon's wave as it radiates from a stellar source is the possibility that this phenomenon could account for an expansion of space around each star due to the untold number of photons emitted over time. Layzer(9) performs the calculation and based on the equivalence energy = momentum = $h\nu$ for photons says that "the frequency of a photon in an expanding universe continually diminishes." Cause and effect could just as logically be reversed to assert that the diminishing frequency of the photon continually expands the universe. If the space around each star were stretched due to its radiation then all stars would appear to be receding for all neighboring stars. The model first proposed by Eddington as dots on an expanding balloon in a two dimensional model of curved space would seem to have a better model of a bowl of exploding popcorn in ordinary three dimensional space. By equating the generation of space to the infusion of energy we are able to propose an additional term to be incorporated into the First Law of Thermodynamics. As Joule first proposed that heat was equivalent to work and expanded the law for the conservation of energy to account for heat, this model suggests that the incorporation of the cosmological constant given by Hubble's constant should likewise be incorporated into the First Law to account for space. Hubble's constant as given by Layzer(10) is defined as: $H = (1/\alpha)(d\alpha/dt)$ where $\alpha$ is the scale factor dependent on time and $r = \alpha(t) r_o$ for any radial direction in Cartesian space at some later time $t$ relative to a reference distance at some earlier time $t_o$. Layzer goes on to derive an expression based on gravitational attraction that relates the Hubble constant to the mean cosmic density: $H^2 = (8\pi/3) \langle\rho\rangle$. As there is no way to verify either of the unknowns in terms of the other there is no certainty that the mass density is the controlling factor in the expanding universe although gravity clearly plays a role in the geometry of space. The fact that the radiation output of each star is different suggests that the Hubble constant is not a universal constant but varies over the universe. The undoubtedly small magnitude of the effect and the difficulty of detecting the differences between stars is reminiscent of the disbelief encountered by Joule's data. Bent(12) cites Faraday's quote about the disbelief of some of his colleagues in the Royal Society: "because they (the calorimetric measurements) all depended on fractions of a degree of temperature, sometimes very small fractions." It would be an interesting test of the model if sufficient rates of recession of stars in clusters could be correlated to their rates of separation and power output after factoring out the known (Newtonian) celestial dynamics.

One thing is certain about the existence of photons and that is that they are created and destroyed (absorbed) by matter and while they exist they move. Their velocity is dependent on the refractive index of the medium through which they travel. The refractive index is defined as the ratio of velocity in the vacuum (a universal constant) to the velocity in the medium: $\eta = c/v$ with $c = 2.997E10$ cm/sec. The slowing of light in denser medium is an obvious consequence of the electronic structure of the atoms making up the medium interacting with the electromagnetic fields of the light. The refractive index of visible light ranges from 2.1 for diamond to 1.5 for glass to 1.00029 for air. The retarding of the light also causes the direction of the light to change by bending the ray so that the ratio of the distance in the vacuum to that in the medium in unit time equals the refractive index, i.e.

$$\eta = v_o / v_m = (d_o/t) / (d_m/t) = d_o / d_m.$$

If the space around a star is expanding and thus more "compacted" then light from another star traveling past its outer regions should interact as in a denser medium and bend radially inward. This, of course, was the prediction which first validated Einstein's General Theory in 1919. Layzer (11) states that the light which just grazes the sun is bend by 0.85E-5 radians = 1.7 arc-sec. As shown in Figure 4, the triangle with the two solar radii at an angle $\phi$ equal to half the bending angle (a) gives the distance ratio equal to $1/\cos \phi$. Using the first three terms in the power series for the cos of an angle:

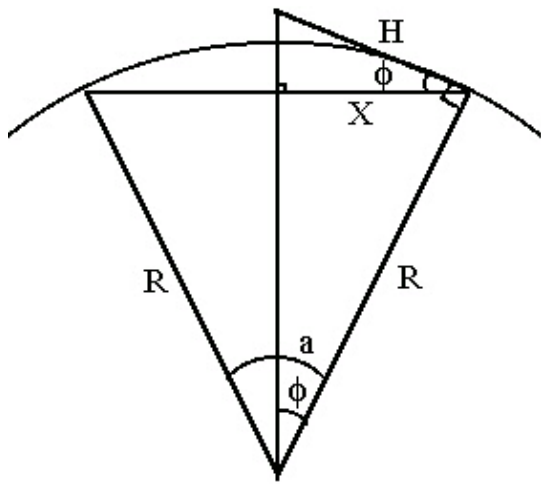

$\cos \phi = 1 - x^2/(2*1) \{= -3.6125E-11\} + x^4/(4*3*2*1) \{= 5.22E-21\}$ it is clear that only the first two terms are significant, and

$$\eta = 1/\cos \phi = 1/(1-3.6125E-11) = 1.000000000031625.$$

Given the phenomena of light bending caused by a massive star, the question of whether space behaves in a material way should be considered a valid inquiry.

Einstein's formulation of the General Field Theory of Relativity relied on the mathematics developed by Riemann who incorporated the concept of space curvature obviously derived from the geometry of spheres as a point of departure. The paradigm of curvature has proven perfectly utilitarian if not confusing when trying to describe curvature in three and four (or higher) dimensions. The point is that what one can model as curvature where it can be imagined can also be modeled identically as compression. Compression is easily imagined in three dimensional space as density gradients and in spacetime as density fluctuations. Three dimensional curvature such as the hilliness of landscapes is commonly represented by two dimensional contour maps. Four dimensional curvature can analogously be represented by three dimensional contours or equivalently by density gradients.

## BLACK-BODY EQUILIBRIUM

One final aspect of universal light phenomena is the existence of residual background isotropic radiation discovered in the mid-twentieth century as microwave noise. This background radiation has been shown to be consistent with the equilibrium radiation known to depend on the temperature of the matter generating it. In the current epoch of the universe this temperature is only around 3 K (colder than the boiling point of liquid helium at atmospheric pressure of 4.2 K and warmer than the liquid helium quantum "lambda" transition of 1.3 K). This equilibrium distribution of radiation exists now without the matter which generated it and is therefore left without a significant means of being altered by reabsorption. The distribution function which describes the equilibrium was first proposed by Planck in 1901 and marked the birth of quantum theory. The function given in differential form is:

$$E\, d\lambda = (hc/\lambda^5)d\lambda / (e^{hc/\lambda kT} -1).$$

For the wavelength in centimeters the units of the right side is (erg sec)(cm/sec)(1/cm$^5$) = erg/cm$^3$/cm. The denominator on the right is unitless. As the differentials have the same units (cm) of wavelength, the energy term is seen to be an energy density. The function has the interesting property that both axes (ordinate and abscissa) can be scaled by the temperature to yield superimposable curves. The graph of $E/T^5$ versus $wT$ is shown in Figure 5 ($\lambda = w$ in the figure) below for plots over temperatures ranging from 10000 K to 1 K. Solar radiation corresponds to a black body temperature of 6000 K corresponding to the range marked by the yellow-green band in the figure for wavelengths in the visible spectrum between $(4000 > 7000) \times 10^{-8}$ cm. (The colored boxes in the figure are shortened on both ends to display the band width, their height has no other significance.) As the temperature decreases the visible band shifts relative to the maximum and the maximum moves to longer and longer wavelengths as the intensity of the curve gets smaller and smaller. Weinberg[13] points out that the wavelength maximum is $\lambda = 0.2014052 \, hc/kT$ or $\lambda T = 0.2014052 \, hc/k = (0.2014052)(1.4389) = 0.2898052$. At a background temperature of this era of 2.7 K, the wavelength of the maximum is 0.1073 cm (0.2898/2.7) and intensity of 4.04E-27 (5.79E-25/$2.7^5$). Twenty-one experimental points compared at 2.7 K by Davies[14] in a log intensity vs. log frequency plot of "several observations of the cosmic background radiation" show good agreement when interpolated at a resolution of 1 part in 500 with a wavelength maximum at 0.172 cm compared to the Planck maximum at 0.111 cm although there is a slightly sharper fit at 2.6 K than at 2.7 K.

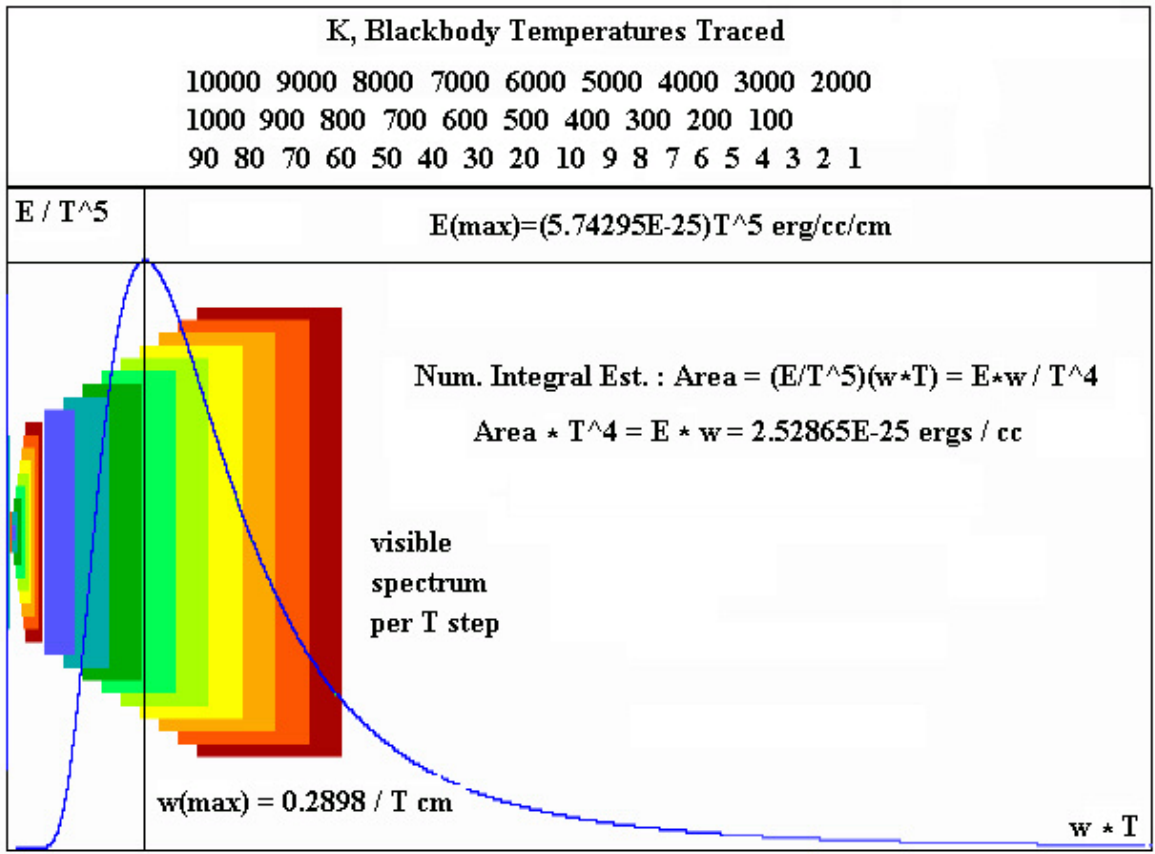

One very interesting aspect of the figure is the integral of the function, i.e. area under the curves, which was estimated numerically. The sum of the products of the intensity at each step in the calculation times the step width over each temperature curve times the temperature of the curve is a constant (within six significant figures but rounded off to four figures) and equals 2.529E-25 erg/cc (cc in the figure is cubic cm, i.e. $cm^3$). To get some feel for the magnitude of the energy density, conversion to mass density ($m = E/c^2$) and atomic density (H atom mass of 1.66E-24 gm) yields one hydrogen atom every 180 km (110 miles) in each perpendicular direction or 3700 atoms in the volume of the moon (2.2 million $km^3$).

Thus for a non-expanding space the energy density is a constant as the radiation cools, which is to say that the energy is conserved and no work was done on cooling. The cooling is caused by the adiabatic (no heat/energy loss) expansion of the space containing the radiation. Adiabatic expansions occur either explosively where the loss of heat is too slow to be transferred out of the volume or when the volume is in isolation and there is no boundary to cross. The system would therefore exhibit no increase in the entropy (disordering) of the space where typically the entropy increase in an expansion is related to the relative change in volume and the relative change in temperature, for initial and final states: $V_f - V_i / V_f = \Delta V/V = \Delta(\log V)$ and similarly $\Delta(\log T)$ as:

$$\Delta S = c_v \ln (V_f/V_i) + k \ln (T_f/T_i) = 0 \text{ and } c_v \ln (V_f/V_i) = k \ln (T_i/T_f)$$

where the T term is inverted to reflect the sign change on equating the two terms. As the atomic heat capacity of an ideal gas is of that of non-interacting particles and equals $3k/2$ then treating the photons as an ideal gas and dividing, taking the antilogs and cross multiplying we are left with the condition:

$V_f^{3/2} T_f = V_i^{3/2} T_i$ or $V^{3/2} T$ is a constant.

## RATES OF EXPANSION

The current epoch is believed to have begun 300,000 years after the Big Bang when the temperature dropped to 3000 K and the interaction between matter and radiation became deactivated. At this point the spontaneous creation of particles from the radiation field was quenched because the energy had decreased below a critical threshold. This energy threshold is given by kT or $1.38E-16 \times 3000 = 4.14E-13$ erg per particle. As this transition was crossed the space in the Universe became transparent as the plasma (proton(+) and electron(-) gas) filling the space condensed into neutral atoms. Given an age for the Universe of 14.8E9 years, the remaining time has consisted in the formation of individual stars, galaxies, and clusters of galaxies acting under the influence of gravity, the only other long range force besides the electromagnetic force, as the Universe expanded. For a present radiant temperature of 3 K, given that $V^3T^2$ is a constant, the temperature decrease from 3000 to 3 (thousandfold) would correspond to a volume increase of a hundredfold. Expansion of the Universe beyond this hundredfold increase in this epoch must be attributable to other factors, such as the creation of space from stellar radiative dissipation.

Within the first minute the Universe had expanded and cooled from a $10^{-8}$ cm and $10^{32}$ K to 4 light years and 520 million degrees and proton, neutrons and electrons were the major constituents of matter awash in a sea of photons. Within the first $10^{-32}$ seconds space (and time) were created and according to an "inflationary" phase transition at $10^{27}$ K that rationalizes several problems of the grand unification theory (GUT) of three of the four fundamental forces the Universe had expanded to about one meter. These figures are based on data presented in a collection of Scientific American reprints from the 1980's(15). From a set of three log-log graphs presented by A. H. Guth and P. J. Steinhardt(16) the dependence of temperature (T, K), energy density (E, ergs/cc), and radius (R, cm) are straight lines plotted versus time having the slopes (d log Y/d log t) of - 0.524, - 2.048, and 0.489 respectively. The time abscissa ranges from $10^{-45}$ (less than the Planck time of $10^{-43}$) to $10^{18}$ seconds (over 15 billion years) in powers of 10. By taking ratios of the three slopes the time base can be eliminated and yields the following interesting ratios: $\Delta \log E/\Delta \log T = 3.91 = \sim +4$, $\Delta \log R/\Delta \log T = - 0.93 = \sim -1$, and $\Delta \log E/\Delta \log R = -4.12 = \sim -4$. Given that $\Delta \log Y/\Delta \log X = q$ can be rearranged to $\Delta \log Y = q \Delta \log X = \Delta \log X^q$, we can deduce that the phenomenal dependences are $\Delta \log E = 4 \Delta \log T$, $\Delta \log R = -\Delta \log T$, and $\Delta \log E = -4 \Delta \log R$. As $d(\log X) = dX / X$, the rate of change is relative, e.g. percentage change, since the amount of change depends on the current (instantaneous) magnitude of the property. These can be compared to those obtained above. For the adiabatic expansion, $V^3T^2$ = constant and converting $V=R^3$ gives $(R^3)^3 T^2 = R^9T^3$ or $R^{4.5}T$ = constant or $4.5d (\log R) = - d (\log T)$ and $d (\log R)/d (\log T) = -4.5$. For both of the black body relations $\lambda T = 0.2898$ = constant and Planck's law: $E = (hc/\lambda^5 T) / (e^{hc/\lambda kT} -1)$, if $\lambda \sim R$ then $RT$ = constant and $d(RT) = R\, dT + T\, dR = 0$ or $dR/R = -dT/T$ and $d \log R = -d \log T$ or $d(\log R)/d(\log T) = -1$ and using $\lambda T$=constant = b in Planck's law then $E\lambda^5 = hc/(e^{hc/kb} -1)$ = constant. Therefore if $\lambda = R$ then $d (ER^5) = R^5 dE + 5ER^4 dR = 0$ and $d \log E / d \log R = -5$ instead of $-4$. These are summarized in the Table.

## COMPARISON OF EXPANSION RATES

| Big Bang Rates: | Big Bang ratios: | Derived Rates | Condition |
|---|---|---|---|
| d log T / d log t = - 0.5 | d log E / d log T = + 4.0 | ----- | ----- |
| d log E / d log t = - 2.0 | d log E / d log R = - 4.0 | d log E / d log R = - 5.0 | $E\lambda^5$ =const. |
| d log R / d log t = + 0.5 | d log R / d log T = - 1.0 | d log R / d log T = - 1.0 | $\lambda T$ = const |
| ----- | ----- | d log R / d log T = - 0.22 | $V^3T^2$ = const. |

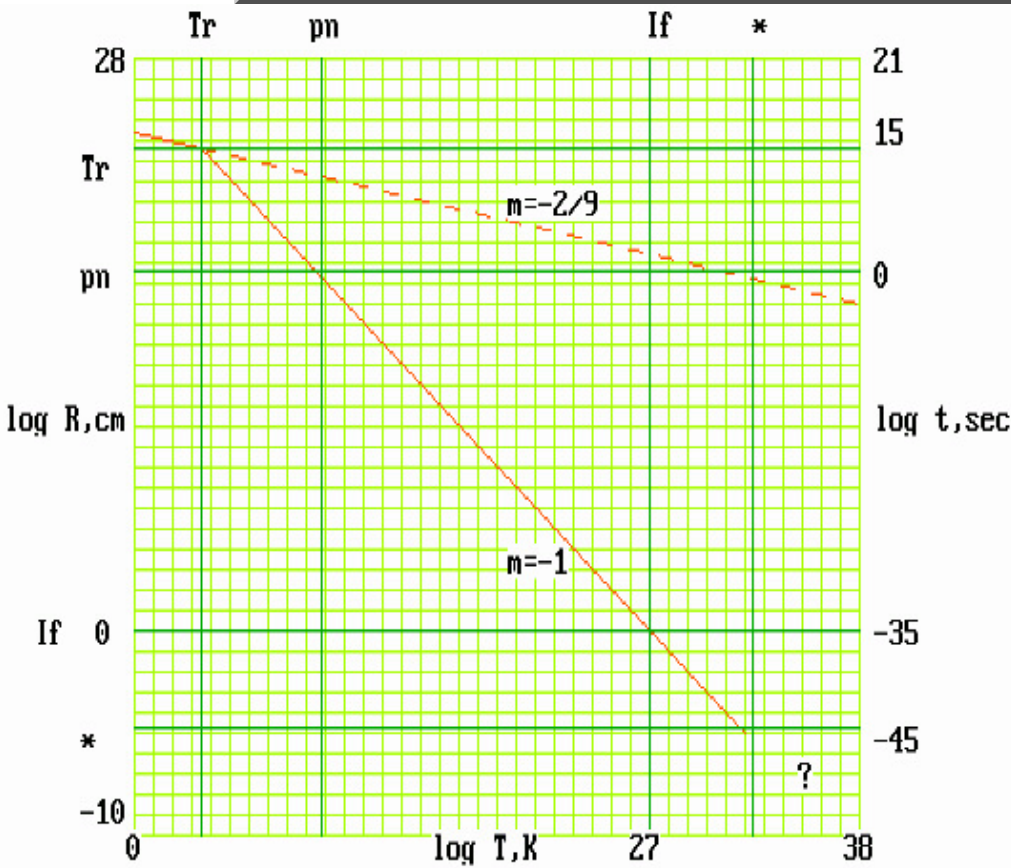

Without knowing how the rates for the Big Bang model were determined it is difficult to make a case against the E vs. R values but it is clear that the black body results agree in the R vs. T rate although that rate precedes the current epoch when matter and light decoupled and the Universe became transparent. These rates are illustrated in Figure 6 as a log R-log T plot and log of time in seconds parallel to the radius coordinate but at half the scale, i. e. five blocks equal 10 powers of 10. The adiabatic expansion rate of 2/9 (=0.22 ) in the graph and would apply only into the Tr (transparency) era of the last 1.0E13 seconds or 300,000 years after the Big Band or 14.5 billion years ago. Clearly the log scale hides a long time span and enlarges a very short span up to 1 second or 1.0E0.

## SPACE

If one reads the literature on space it is hard to find any source that doesn't start with geometry and takes as given the assumed nature of space. Clearly, the common philosophical (psychological?) distinction of space is concerned with the relative positioning of objects with respect to each other. In the Forward to Jammer's treatise, Einstein[17] provided a succinct summary of that nature: "These two concepts of space may be contrasted as follows: (a) space as positional quality of the world of material objects; (b) space as container of all material objects. In case (a), space without a material object is inconceivable. In case (b), a material object can only be conceived as existing in space; then space appears as a reality which in a certain sense is superior to the material world. . . . These schematic considerations concern the nature of space from the geometric and from the kinematic point of view respectively." Historically, the nature of space has evolved through three eras that can be encapsulated by three individuals: Euclid, Newton, and Riemann. Of course, the mathematics of geometry was developed early in recorded history and Euclid's elaboration of its logical structure of definitions, axioms, propositions and proofs is well known even by some high school students. With the birth of classical physics the turmoil between Newton's concept of absolute space and light corpuscles opposed by Huygens and Leibniz relativism and wave theory in which he won one and lost one only to have those fortunes reversed after a two hundred year interlude. The third era was the discovery of non-Euclidian geometry beginning with Gauss and ending with Einstein. Both the Special theory and the General Field theories had created need for mathematical treatments of the structure of space. In summarizing Riemann's contributions, Jammer[18] noted: "Once again we see that, historically viewed, abstract theories of space owe their existence to the practice of geodetic work, just as ancient geometry originated in the practical need of land surveying."

A more recent description of the positional concept was made by Smolin [19] in which he first states "space is nothing apart from the things that exist"(p.18) but that it is not necessary for material things to exist because there are fields that can exist in space without matter ("material particles"). In particular, he cites electromagnetic and gravitational fields, but notes that these fields "vary continuously over space." He concludes his discussion of space-time by stating that "Time is nothing but a measure of change - it has no other meaning." It would appear that Smolin has simply substituted the word "field" for the meaning of the word "space." With the dismissal of time as a dimension, it seems that the special relativistic concept of space-time is to be relegated in the same way that "aether" was relegated because of its absolutist flaw. There are probably no volumes of the universe that are field-free but they are phenomenal volumes nonetheless however uncertain their geometry and in which two observers could in principle move at different velocities and observe differences in simultaneity.

The opposite treatment of space, i.e. as substantive, is not new. Of particular relevance to our discussion is the concept of aether which was laid to rest by the famous Michaelson-Morley experiment around the turn of the 20th century. It was to be replaced by the concepts of macroscopic "space" and microscopic "vacuum." The aether had been postulated to provide the medium through which electromagnetic waves were propagated (much as air is the medium for sound). More significantly, it was also thought to be an absolute rest frame. It was this latter property which was it's fatal flaw. Unfortunately the two substitutional terms carry as much or more connotational baggage as "aether." Particle physicists and quantum mechanicians typically refer to the space existent at their level of interest as the "vacuum." This concept is of space devoid of matter but serving as a physical medium for interactions and a fluctuating source of "virtual" particles. In the Danz lecture series, Lee[20] suggested: "This difficulty could be resolved by introducing a new element, the vacuum. Instead of saying that the symmetry of all matter is being violated, we suggest that all conservation laws must take both matter and vacuum into account. In this case, Lee's position corresponds to Einstein's example of kinematic space.

The treatment of space as substantive was presented around 1870 by the mathematician William Kingdon Clifford[21] in a brief reaction to Riemann's work prior to the quantum or relativity revolutions of the early 1900's. He proposed four speculations: "(1) That small portions of space are in fact of a nature analogous to little hills on a surface which is on the average flat; namely that the ordinary laws of geometry are not valid in them. (2) That this property of being curved or distorted is continually being passed on from one portion of space to another after the manner of a wave. (3) That this variation of the curvature of space is what really happens in that phenomenon which we call motion of matter, whether ponderable or etherial. (4) That in the physical world nothing else takes place but this variation, subject (possibly) to the law of continuity."

Three further explicit statements of this nature all date to the mid-1980's. The first was verbal rather than written and spoken by the astronomer and student of Hubble, Allan Sandage[22], who made the following remark in the 1985 PBS television program 'The Creation of the Universe': "It is not as if these galaxies are expanding into a space that's already there. The view is that space itself is expanding carrying the galaxies with it. The expansion creates the space." Harvard astrophysicist, David Layzer[23] writes: "The cosmic expansion is an expansion of space itself, not a systematic recession of galaxies in a static space." P. C. W. Davies[24], a professor of theoretical physics, wrote"The expansion of the universe, discovered by Edwin Hubble in the 1920s, is a cornerstone of modern cosmology, and is best envisaged as the continual swelling or stretch of space itself." Careful reading of any of these statements suggests that space is not being created but that it is "infinitely(?)" elastic. One wonders why in a science so imbued with notions of conservation is space generation so freely allowed to come from nothing.

## SPACETIME

Little mention is given of what was involved at the moment of creation. Davies[24] gives the one of the best statements: "Cosmologists believe that the big bang represents not just the appearance of matter and energy in a pre-existing void, but the creation of time and space too. The universe was not created in space and time; space and time are part of the created universe." The question is how space and time were brought about from the supposed singularity. Realizing that the singularity was a point of immense energy content and that the uncertainty principle requires that the precision of either of the conjugate pairs: momentum and position or energy and time must satisfy Heisenberg's condition of $\Delta p \Delta x > h/2\pi$ or $\Delta E \Delta t > h/2\pi$. There is phenomenon in spectroscopy known as lifetime broadening which relates the frequency width of an energy state of matter, i.e. the precision in specifying the quantum energy state. It is a consequence of the solution of the time-dependent Schroedinger equation and has the same form as the uncertainty principle where the broadening of the energy is related to the lifetime (stability) of the state.(25) Since there can be no infinite precision, it seems plausible that there was within the uncertainty limits of the lifetime of the energy the creation of a one-dimensional position in space at least the length of the uncertainty of the position.

## ONE DIMENSION.

There is sufficient evidence of the utility of associating an imaginary orthogonal axis to any real axis to warrant the faith of its (curious choice of terms) reality. An enlightening example is described by G. 't Hooft(26) in reference to an electron's behavior as a wave in an electric field: "The electron field is a moving packet of waves, which are oscillations in the amplitudes of the real and imaginary components of the field." The physical world has abundant evidence of phenomenal functions containing the imaginary quantity, $i$. Solutions of Schroedinger's equation ($H\Psi = E\Psi$) for the hydrogen atom allows separation of the spherical variables: $\Psi(r, \theta, \phi) = R(r) \Theta(\theta) \Phi(\phi)$. The first two factors are Real and dependent on the quantum numbers n (radial) and l (longitudinal) while the third which is a function of the magnetic quantum number m (latitudinal) has solutions of the form $e^{im\phi}$. Because of their imaginary dependence, physical solutions of the m equations have to be taken as linear combinations.

In general, numerical quantities can be expressed in the complex form:

$$n = a + bi,$$

where $i$ is the square root of minus one and a and b are real numbers. Real numbers have b=0 and purely imaginary numbers have a=0. It is possible to bring some insight into the mathematics of complex numbers using two dimensional plots of orthogonal axes representing the Real and Imaginary components. These are the so-called Argand diagrams detailed in Figure 7.

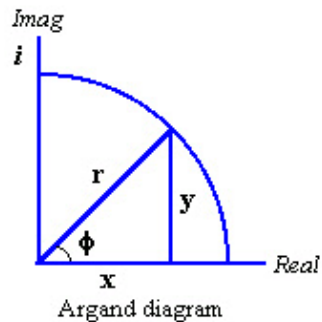

$x = r \cos \phi$
$y = r \sin \phi$
$x^2 + y^2 = r^2$
$e^{\pm i\phi} = \cos \phi \pm i \sin \phi$
$= x/r \pm iy/r$
$re^{\pm i\phi} = x \pm iy$

Argand diagram
**One-dimensional Plane Intersection of Real and Imaginary Axes**

## TWO DIMENSIONS.

If we consider two perpendicular planes and label one Real and the other Imaginary analogous to the two dimensional axes of an Argand diagram, then a three dimensional representation of the exponential function $e^{i\phi}$ given by Gauss's equation: $e^{i\phi} = \cos \phi + i \sin \phi$ shown in Figure 8 is obtained from the following vector algebra:

| TERM | RADIUS | FUNCTION |
|---|---|---|
| $x_{imag} = r \cos \phi$ | $r = \sin \phi$ | $x_{imag} = \cos \phi \sin \phi$ |
| $y_{imag} = r \sin \phi$ | $r = \sin \phi$ | $x_{imag} = \sin^2 \phi$ |
| $x_{real} = r \cos \phi$ | $r = \cos \phi$ | $x_{real} = \cos^2 \phi$ |
| $y_{real} = r \sin \phi$ | $r = \cos \phi$ | $y_{real} = \sin \phi \cos \phi$ |

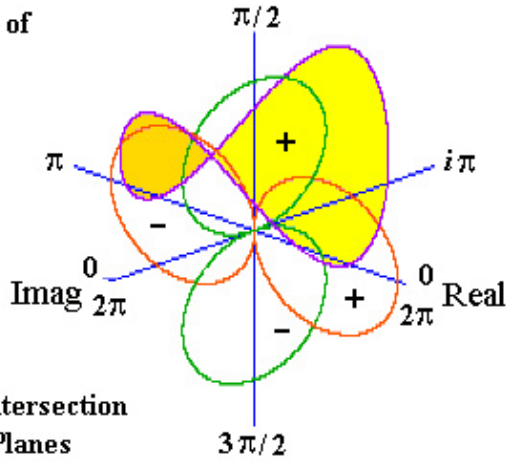
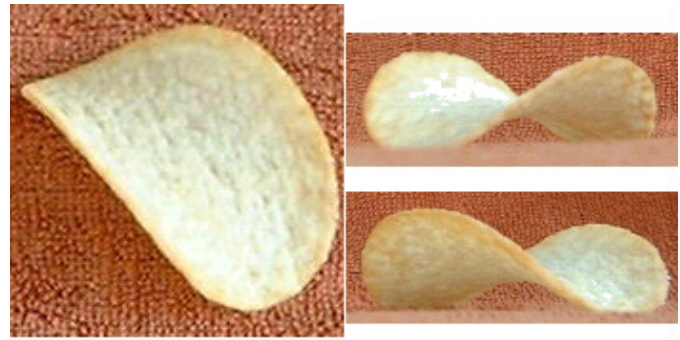

Plotting the Real radius vector, $r_{real} = x_{real} + y_{real}$ as a function of $\phi$ in the Real plane and the Imaginary radius similarly in the Imaginary plane and then adding these two radius vectors as a function of $\phi$ yields the plane bounded at unit radius representing $e^{i\phi} = \cos^2\phi + 2\cos\phi\sin\phi + \sin^2\phi = (\cos\phi + \sin\phi)^2 = 1 + \sin 2\phi$. A student named it the pringle function after the potato chip of the same shape shown in Figure 9. Projections of the two orthogonal views of the "chips" in the figure onto the real plane correspond to the hydrogen-like atomic orbitals labeled $p_x$ and $p_y$ for the m quantum number equal to +1 and -1 which would be obtained as linear combinations. The $p_z$ orbital has m=0 and therefore has no imaginary component. The midlines of the curve align with the orthogonal axes. The minima and maxima occur along the 45º lines in each of the quadrants: minima values of 0 (zero) at 135º and 315º and maxima values of 1 (unity) at 45º and 225º. One of these latter points illustrates the fascinating relation:

$$e^{i\pi} - 1 = 0;$$

the other point illustrates that any base (here e) raised to the zeroth power is unity.

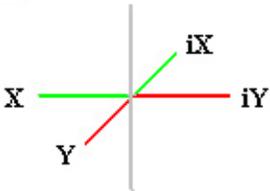

THREE DIMENSIONS.

Extending the so-called Argand diagram to three dimensional space, we assume that there exist three real orthogonal dimensions, viz. X, Y, and Z in Cartesian (Euclidian) space and also assume that there exists an orthogonal imaginary axis to each real axis, e.g. $i$X, $i$Y, and $i$Z. In essence this follows Hamilton's quaternion definition(27): $\mathbf{q} = t + u\mathbf{i} + v\mathbf{j} + w\mathbf{k}$, where his single real dimension is split into three: $t = a\mathbf{x} + b\mathbf{y} + c\mathbf{z}$ with X = $a\mathbf{x} + u\mathbf{i}$ etc. It is not relevant to define the negatives of any of these six axes because the position of an origin is relative and negative values simply imply a change in direction along an axis. The most symmetrical way to combine these six axes is to place two pairs of them in a plane such that, say X vs. $i$X and Y vs. $i$Y, are at right angles with $i$X as the linear extension of Y and $i$Y as that of X differing from the two-dimensional case described above. To place the remaining real axis, say Z, perpendicular to the X-Y plane, we find that for the $i$Z axis to be orthogonal to the other axes it must coincide with one of the four in the plane. Whatever the mechanism for the coincidence, there exists the opportunity for it ($i$Z) to interact with one of the other imaginary axes in the plane, say $i$X, to produce the cross product and become real and negative ($i^2$ XxZ = -Y). That this product takes the "Y" label is a consequence the naming system. As the resultant axis must be orthogonal to the plane (XY) of its multipliers the Y labeling must be an artifact. Moreover, the "negative" value of the product dimension is exceptional for a system where the "negative" direction is also assumed to be artificial. There are two options for interpretation at this point. We can either reject the generation of the "minus Y" dimension because of these two contradictions or we can accept the generation of a fourth real negative axis orthogonal to the other three real axes. It would seem that since we know that such a system exists phenomenally it would behoove us to accept the latter option. Since the labeling of the various Cartesian axes is arbitrary, we are left with five mutually orthogonal dimensions: three positive real, one imaginary, and one negative real dimension: x, y, z, $i$, and t. It would seem that the fourth real (negative) dimension clearly is time.

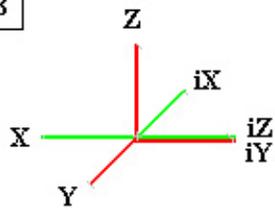

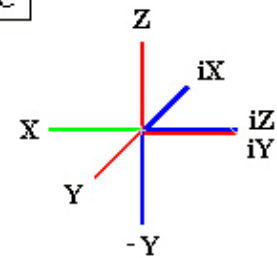

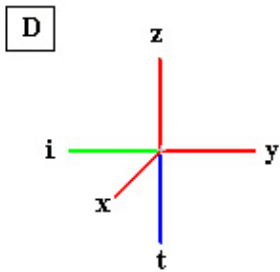

D

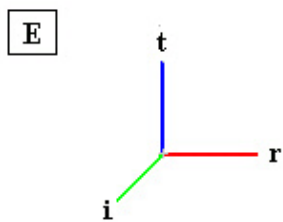

E

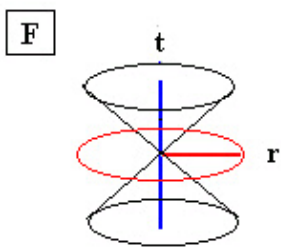

F

Given the phenomenal correlations associated with wave phase relations and the imaginary dimension, the generation of a time axis from imaginary dimensions has an esthetic appeal. The fact that there is only one imaginary axis when there are three real ones is not really a concern. The linear translation of a mass point reduces the three Cartesian axes to a single radial distance:

$$r = (x^2 + y^2 + z^2)^{1/2},$$

orthogonal to the single imaginary axis, *i*, in the standard two dimensional Argand representation. Similarly, plots of Minkowski's space-time geometry are as readily described in a three dimensional representation using the time ordinate and an orthogonal polar plane with a radial distance to describe the light cone and regions of time-like and space-like events. Once again because the labeling is arbitrary, we can reverse the polarities of the Cartesian dimensions with the time dimension in the evaluation of the Minkowski-Pythagorean (proper) interval:

$$s^2 = t^2 - (x/c)^2 - (y/c)^2 - (z/c)^2.$$

where now the minus signs do not affect the implicit positive distances.

## REFERENCES AND NOTES

1. Peter Woit, Not Even Wrong, Basic Books, 2006, p.256
2. Lee Smolin, The Trouble With Physics, Mariner Books, 2007, p.256
3. P. A. M. Dirac, The Mathematical Foundations of Quantum Theory, Mathematical Foundations of Quantum Theory, A. R. Marlow, Ed., Academic Press, 1978, pp.1-8
4. G. N. Lewis and M. Randall, Thermodynamics, Rev. by K. S. Pitzer and L. Brewer, McGraw-Hill, 1961, p.23
5. Richard C. Tolman, Relativity, Thermodynamics, and Cosmology, Clarendon Press, 1934; Dover, 1987, pp.7-9
6. Joel H. Hildebrand, John M. Prausnitz and Robert L. Scott, Regular and Related Solutions, Van Nostrand Reinhold Co., 1970, p.vi
7. J. Schwartz, The Pernicious Influence of Mathematics on Science, Proc. of the 1960 International Congress, Stanford University Press, 1962
8. cf. Mary L. Boas, Mathematical Methods in the Physical Sciences, 2nd Ed., John Wiley & Sons, 1983, Chapt. 1
9. David Layzer, Constructing the Universe, Scientific American Books, 1984., p.239
10. Layzer, op. cit., p.227 ff.
11. Henry Bent, The Second Law, Oxford University Press, 1965, p.15
12. Layzer, op.cit., p.212
13. Steven Weinberg, The First Three Minutes, Basic Books, 1988, p.174
14. P. C. W. Davies, The Accidental Universe, Cambridge University Press, 1982, p.32
15. John D. Barrow and Joseph Silk, The Structure of the Early Universe, Particle Physics in the Cosmos, R.A.Carrigan and W. P. Trower, Eds., W. H. Freeman, 1989, pp. 22-36
16. Alan H. Guth and Paul J. Steinhardt, The Inflationary Universe, op.cit., p.186
17. Max Jammer, Concepts of Space, 2nd. Ed., Harvard University Press, 1969, p.xiii
18. op.cit. p. 152
19. Lee Smolin, Three Roads to Quantum Gravity, Basic Books, 2001, p.18-24
20. T. D. Lee, Symmetries, Asymmetries, and the World of Particles, University of Washington Press, Seattle, 1988, p.47
21. William Kingdon Clifford, Mathematical Papers by William Kingdon Clifford, Ed. Robert Tucker, Chelsea Pub. Co., reprinted 1968, pp. 21-22; cit. Cambridge Phil. Soc. Proc., II, 1876. Read Feb. 21, 1870, pp.157-158.
22. Allan Sandage, 1985 PBS television program 'The Creation of the Universe'
23. Layzer, op. cit., p.244
24. P. C. W. Davies, The Accidental Universe , Cambridge University Press, 1983, p. 4
25. Paul Davies, The Cosmic Blueprint, Simon and Schuster, 1988, p.123

26. cf. Peter Atkins, Physical Chemistry, 5th Ed., W. H. Freeman & Co., 1994, p.553
27. Gerald t'Hooft, Gauge Theories of the Forces between Elementary Particles, Particle Physics in the Cosmos, R.A.Carrigan and W. P. Trower, Eds., W. H. Freeman, 1989, p.84
28. cf. Roger Penrose, The Road to Reality, Vintage Books, New York, 2007, Chapt. 11

[1] Department of Chemistry, Virginia Polytechnic Institute and State University, Blacksburg, VA 24061; Internet: pfield@vt.edu.
Paul E. Field received his Ph.D. in physical chemistry from Penn State in 1963 at which time he joined the faculty at Virginia Tech. He taught in the Chemistry Department until he retired in 2000. His research interests were in the thermodynamics of the liquid state. His teaching interests were in physical chemistry, chemical bonding, thermodynamics, and computer programming and interfacing for laboratory automation. He has had particular interest in cosmology for the past twenty-five years.

[2]The series $0.414 \sin(2X) + 0.412e{-1} \sin(4X) + 0.414e{-3} \sin(6X)$ gives a least squares fit between $-3\pi$ and $+3\pi$ of $4.76e{-4}$ for the summed exponentials; N.B. this excludes the exponential trough and crest just beyond the limits.